\documentclass{zermatt}
\usepackage{graphicx}
\usepackage{natbib}
\usepackage{amssymb}
\usepackage[T1]{fontenc}
\usepackage[utf8]{inputenc}
\usepackage[english]{babel}
\usepackage[scaled]{helvet}
\begin{document}
\title{Star Formation in the Extreme Galactic Center Environment} 
\runningtitle{Galactic Center Star Formation}
\author{Mark R. Morris}
\address{Department of Physics \& Astronomy, University of California, Los Angeles, CA, USA 90095-1547\\ \email{morris@astro.ucla.edu}}
\begin{abstract}
Copious star formation occurs in the dense Central Molecular Zone (CMZ) of our Galaxy, but at a much smaller rate than occurs in a comparable mass of molecular gas in the Galactic disk.  The combination of large turbulent velocity dispersions, a relatively strong magnetic field, and a strong tidal field all contribute to inhibiting star formation (SF) in different ways in different CMZ locations.  Nonetheless, there are spectacular displays of recent and ongoing SF in the CMZ, including massive young stellar clusters, sites of abundant SF in progress, and numerous spots of protostellar or YSO activity.   The presence of giant molecular clouds in the CMZ that are almost entirely devoid of SF indicates that SF requires a trigger that is not present everywhere.  The dominant provocation of SF is likely to be cloud compression, either by large-scale shocks or by orbital motion of clouds into a region of enhanced tidal compression and/or enhanced external pressure.   Recent hypotheses for where and how SF takes place in the CMZ are constrained by the recent orbital determinations of the massive Arches and Quintuplet clusters.  Star formation in the central parsec is subject to a very different set of physical conditions, and is less well understood, but is important for the co-evolution of the central black hole and the nuclear star cluster.   

\end{abstract}
\maketitle
\section{Introduction}
\label{Morris:sec:intro}

 The discussion of SF in the Galactic center divides naturally into that taking place in the gravitational domain of the Galactic black hole (GBH, i.e., roughly within a parsec) and that taking place in the rest of the CMZ.  The very strong tidal forces exerted by the GBH in the central parsec raise the threshold for SF considerably, and it remains debatable whether SF has occurred very recently there, but the presence of the young nuclear cluster within half a parsec of the GBH shows that a dramatic SF event did occur several million years ago, unavoidably accompanied by a major episode of accretion onto the GBH.  Here, the two regimes -- CMZ and central parsec -- are discussed separately.  For complementary recent overviews of star formation in the CMZ, the reader is referred to \citet{Krumholz20} and \citet{Henshaw+22}.

\section{The Large-Scale Arena: the Central Molecular Zone}
\label{Morris:sec:CMZ}

 The CMZ contains on the order of $5\times10^7$ M$_{\odot}$ of molecular gas lying within $\sim\pm200$ pc of the Galactic center \citep{Dahmen+98,Ferriere+07,Longmore+13a}, the majority of which is organized into a coherent structure, or structures, surrounding the nucleus.  That structure might be be a twisted, elongated closed ring \citep{Molinari+11}, several gas streams following open orbits \citep{Kruijssen+15,Henshaw+16}, or relatively tightly wound spiral "arms" of gas \citep{Sofue95a} that more or less coincide with the most prominent orbiting streams.  In any case, molecular "clouds" in the CMZ tend to be denser portions of tidally stretched molecular gas streams or of a ring \citep{Molinari+11, Butterfield+18,Kruijssen+19}.  Simulations show that closed rings of dense gas form naturally in the presence of a bar potential \citep{KimSS+11,KrumholzKruijssen15,Sormani+18b,Salas+19, Sormani+20a,Tress+20}, generally occupying the region of the outermost X2 orbits \citep[c.f.,][]{MorrisSerabyn96}.  
 
 Because most of the dense CMZ mass is in the $\sim100$ pc ring (or the component streams of the ring distribution), one should expect Galactic center star formation to predominantly take place there.  However, although the star formation rate is substantial \citep[$\sim0.1$ M$_{\odot}$ yr$^{-1}$; c.f.,][and references therein]{Barnes+17,Henshaw+22}, the rate of star formation per unit mass of gas is much smaller in the CMZ than in the Galactic disk \citep{Morris89_IAU,Longmore+13a, Longmore14b}.  The dominant reason for the inhibition of star formation in the CMZ is likely to be the large energy density of turbulence, or microturbulence, which is implied by the large linewidths observed with even the highest possible spatial resolution (typically $\sim10$ km s$^{-1}$ or greater). This is attributable to several causes: supernovae and other feedback from star formation, supersonic turbulence induced by MHD instabilities in the differentially rotating medium, and by instabilities accompanying bar-driven angular momentum transport \citep{KrumholzKruijssen15}.  The large velocity dispersions associated with turbulence effectively raise the Jeans mass to values akin to the masses of clusters rather than individual stars, which could be relevant to the formation of the massive Arches and Quintuplet clusters in the CMZ.  
 Another contributing reason for the relatively low rate of SF is the relatively strong magnetic field in clouds of the CMZ \citep{Morris93,Chuss+03}, which provides a  pressure that anisotropically counteracts gravitational collapse. The magnetic field in CMZ clouds is likely to be closely linked to, and in energy equipartition with, the turbulence.  Galactic tidal shear has been considered as another factor inhibiting star formation, but its effect is likely to be negligible except in the central parsec \citep{Kruijssen+14}.
 
 The supersonic turbulence within CMZ clouds must inevitably produce an unresolved network of shocks. Indeed, observations of molecular shock tracers indicate that shocks are present throughout the CMZ.  Those tracers include  SiO \citep[][and Riquelme, this conference]{Riquelme+10A,Tsuboi+11}, HNCO \citep{Henshaw+16}, and hot NH$_3$ \citep{MillsMorris13}.  Density enhancements and cooling in post-shock regions might lead to conditions favoring star formation, but given the strong magnetic fields in CMZ clouds, the shocks are likely to be of type C, so these effects are probably somewhat muted. In any case, there is currently no evidence for a correlation between the locations of shocks and sites of star formation.  Indeed, the internal shocks in CMZ clouds undoubtedly play a strong role in maintaining a relatively high gas temperature in the molecular gas \citep{Immer+16}, which contributes to the impediments to star formation.  
 
 An important environmental factor affecting star formation in the CMZ is that the pressure there, including turbulent gas pressure and magnetic pressure, $P/k\sim10^8$ K cm$^{-3}$, is several orders of magnitude higher than in the Galactic disk \citep{SpergelBlitz92,Rathborne+14b}.  According to the analysis of Rathborne et al., this translates into much higher surface densities of CMZ clouds compared to Galactic disk clouds, and therefore much higher volume densities to be in hydrostatic equilibrium with the hydrostatic pressure from self-gravity.  Indeed, \citet{Rathborne+14b} find that the physical conditions and structures of CMZ clouds are close to those representing hydrostatic equilibrium, which accounts for their relatively high densities throughout the clouds, typically $\gtrsim 10^4$ cm$^{-3}$.

 \section{Where and Why Does Star Formation Take Place in the CMZ?}
 \label{Morris:sec:where}
 
There are a few places in the CMZ where star formation has apparently been provoked by some local compressive event.  An example is G-0.02-0.07, a linear string of 4 HII regions lying adjacent to a dense ridge in the 50 km s$^{-1}$ molecular cloud \citep{Serabyn+92,YZ+10b,Mills+11}. 

However, there are two hypotheses for how most of the star formation in the CMZ is initiated in a more global manner.  The first hypothesis, which has been referred to as the "conveyor belt" model \citep{Longmore+13b,Longmore+14a, Kruijssen+15, Henshaw+16b,Kruijssen+19}, notes that the trajectories of the streams and their component clouds constituting the molecular ring are elongated, so that the clouds pass through a pericenter position where they orbit closest to the GBH.  At that point, the clouds are maximally compressed by the Galactic tidal force imposed by the GBH plus the nuclear stellar cluster, and at least portions of those clouds are pushed over the threshold of gravitational stability.  
The proponents of this hypothesis point out that clouds distributed along the gas streams following passage through the pericenter location display a sequence of evolutionary stages for star formation, from clouds that are almost entirely devoid of star formation (e.g., G0.253+0.016, the "Brick"; \citet{Kauffmann+13}), to the massively star-forming cloud, Sgr B2 \citep{Ginsburg+18}.   
\citet{Kruijssen+15} note that the orbital time between the locations of those clouds corresponds to the free-fall time for star formation, once the collapse has been initiated.  Another phenomenon that adds to cloud compression at the pericenter position is the fact that there is probably a radial gradient in the pressure of the interstellar medium, which is greatest near the GBH.

The second hypothesis for large-scale triggering of star formation is that cloud compression occurs where gas flowing inward along the bar from outside the CMZ encounters and shocks the gas in the CMZ, triggering star formation at the {\it apocenters} of the CMZ cloud orbits \citep{Sormani+20a,Tress+20}. In this model, which is based on detailed hydrodynamical simulations, the sequence of star formation begins at the apocenters of the cloud orbits, and the newborn stars are presumably arrayed as "beads on a string" at downstream positions along the X2 orbit ring.  

Assessment of these contrasting hypotheses would be facilitated by having line-of-sight distance measurements or well-determined orbits for recently-formed stars in the CMZ.  The massive, young Arches and Quintuplet clusters can be helpful in this regard.  \citet{Sormani+20a} suggest that these clusters originated near the collision sites where the bar-driven inflow accretes onto the CMZ.  This view is supported by the recent work of \citet{Hosek+22}, who determined the absolute proper motions of these clusters, and combined them with radial velocities and 2D locations to strongly constrain their Galactic orbits.  Although the line-of-sight distances and the precise cluster ages remain undetermined, the constraining information at hand allowed \citet{Hosek+22} to show that the probability maps for the cluster birth locations favor the hypothesis that they originated at the apocenters of the gas ring.  

Numerous studies have revealed the presence of many individual massive stars, HII regions, and embedded protostars spread throughout the CMZ \citep[e.g.,][]{YZ+09,Mauerhan+10a,Mauerhan+10c, An+11,Dong+12, Hankins+19, LuXing+19a, Clark+21}.  \citet{Clark+21} enumerate $\geq 320$ spectroscopically classified stars that would be expected to undergo core collapse within the next $\sim20$ Myr, and state that this number is a substantial underestimate.  Work on the mid-infrared SOFIA survey by \citet{Hankins+20} is continuing and will add considerably to the known population of embedded stars and protostars.  The overall collection of such objects will eventually be invaluable for determining the temporal and 3-dimensional spatial distribution of star formation in the CND.  

Some of the massive stars distributed within 
the CMZ could be escapees from the massive Arches and Quintuplet clusters, distributed along the tidal tails of these clusters  \citep{Mauerhan+10c,Habibi+14}.  In addition to tidal evaporation, supernovae in binaries and 3-body gravitational interactions cause the occasional expulsion of massive stars from such massive, dense clusters \citep{KimSS+99,KimSS+00, PortegiesZwart+02}.  It is therefore possible that a sizeable fraction of the population of isolated, massive field stars could have originally formed in massive clusters.  However, in order to assess this possibility, considerable work remains to determine the dynamics of the isolated stars to see whether their orbits trace back to the clusters.  In any case, the fact that massive clusters contain, or formerly contained, such a significant fraction of all massive stars in the CMZ provides an extremely salient clue to a key mode of star formation there.

\section{The Central Parsec}
\label{Morris:sec:YNC}

The Young Nuclear Cluster (YNC), lying within a radius of 0.5 pc of the GBH, and having an age of 3 - 8 $\times10^6$ years \citep{LuJR+13,LuJR18}, illustrates that star formation is possible within the immediate vicinity of the GBH, despite the extreme tidal force exerted by the GBH, once considered an insurmountable barrier to star formation \citep{Sanders92}.   The possibility that the YNC could have migrated in from elsewhere as a result of dynamical friction has been considered \citep{Gerhard01}, but this would only be possible for a cluster having a dense core that is much more massive than the YNC, even if the cluster is centered on an intermediate-mass black hole \citep{KimSS+03,KimSS+04}.  However, \citet{Fujii+08} found that the inspiral timescale for clusters onto the GBH is reduced by core collapse of massive clusters, so it might have been possible for the YNC to migrate into the center from a distance of a few parsecs within the lifetime of the YNC.  In any case, the prevailing opinion now seems to be that the YNC formed {\it in situ}, but that raises many questions about how local gravitational collapse to form stars could have overcome the extreme tidal shear and probable strong magnetic fields, amplified by the shear, in that region.  Furthermore, formation of a few $\times~10^4$ M$_{\odot}$ cluster that close to the GBH would almost certainly have been accompanied by strong accretion activity that would have added considerably to the heating and turbulence. 

A fraction ($\geq20\%$) of the stars in the YNC form a disk around the GBH, albeit with a relatively large distribution of eccentricities \citep[][and references therein]{LevinBeloborodov03,Yelda+14}.  This suggests that at least some of the star formation leading to the YNC occurred in an accretion disk (although an inspiralling cluster would also leave stars distributed in a disk, once it dynamically relaxed at the center).  A numerical model produced by \citet{Nayakshin+07} showed that a stellar disk could be formed starting from a gravitationally unstable gaseous disk surrounding the GBH. It would be interesting and timely to repeat such a calculation with additional physics (magnetic fields, black hole feedback), and higher spatial resolution to capture the disk instabilities.  The physical conditions in the accretion disk maelstrom that would have produced the YNC are far from ordinary, so it is no surprise that the mass function of the YNC is somewhat top-heavy, with a slope of $1.7\pm0.2$ \citep{Bartko+10,LuJR+13}.  (Note that a top-heavy mass function would also be expected for the surviving members of a cluster that has migrated to the center as a result of tidal friction \citep{Fujii+08}.) 

An alternative model for {\it in situ} production of the YNC that has been very popular invokes an infalling molecular cloud or tidal disruption of a cloud approaching the GBH on a highly eccentric orbit \citep[e.g.][and many others cited by Dinh et al., 2021]{BonnellRice08, Generozov+22}.  However, \citet{Dinh+21} argue that the phase-space volume accessible to such low-angular-momentum clouds is very small, and there is no obvious way of producing them, so they should be quite rare. Put another way, it is very difficult to scatter clouds onto approximately radial orbits because of their typically large size, because they participate in Galactic rotation, and because of the likely absence of sufficiently massive perturbers.  Invoking cloud collisions to produce infalling clouds \citep[e.g.,][]{HobbsNayakshin09} faces similar constraints \citep{Dinh+21}.  Another alternative model for {\it in situ} formation of the YNC is one in which the circumnuclear gas disk, fed quasi-continuously from the outside by gas migrating inward from the CMZ, underwent viscous evolution prior to the formation of the YNC, and extended much further in toward the center than it does now, perhaps all the way to the GBH.  At that point, some trigger, perhaps enhanced pressure from accretion activity, started a runaway process of star formation within the central half parsec of the disk.  In such a circumstance the combined energy produced by the rapid formation of a massive cluster and the near-Eddington accretion onto the GBH could have expelled much of the central portions of the CND, creating a central cavity such as that which is present today \citep{Morris+99}. Noting that accretion onto the CND is likely continuing \citep[e.g.,][]{Tress+20}, and that the lifetimes of the massive stars in the YNC are only a few $\times10^7$ years, one can speculate that the formation of a YNC follows a limit cycle, repetitively recreating such a cluster on a time scale limited by the CND growth rate and by the viscous evolution timescale of the CND.

Evidence for more recent star formation at various places in the central parsec has been presented \citep[e.g.,][]{Eckart+13,YZ+13,YZ+15a,YZ+15b,YZ+17b}, but has been questioned (Morris 2021\footnote{https://www.youtube.com/watch?v=8ezDLs\_MJxw}), largely because there are alternative explanations for the phenomena presented as evidence. Furthermore, observations of the infrared counterparts to the putative protostars and YSOs, or the supporting spectroscopic evidence, have not yet been acquired.  The issue of whether star formation in the central parsec is a continuing and current process will probably soon be resolved by ongoing observations with JWST.

\bibliography{refs}{}

\end{document}